\documentstyle[12pt]{article}
\begin{document}
\centerline{\Large\bf Gravitational Geons in 1+1 Dimensions}
\vskip .7in
\centerline{Dan N. Vollick}
\vskip .1in
\centerline{Irving K. Barber School of Arts and Sciences}
\centerline{University of British Columbia Okanagan}
\centerline{3333 University Way}
\centerline{Kelowna, B.C.}
\centerline{Canada}
\centerline{V1V 1V7}
\vskip .1in
\centerline{and}
\vskip .1in
\centerline{Pacific Institute for Theoretical Physics}
\centerline{Department of Physics and Astronomy}
\centerline{University of British Columbia}
\centerline{6224 Agricultural Road}
\centerline{Vancouver, B.C.}
\centerline{Canada}
\centerline{V6T 1Z1}
\vskip .9in
\centerline{\bf\large Abstract}
\vskip 0.5in
\noindent
It is well known that general relativity does not admit gravitational geons that are stationary, asymptotically flat,
singularity free and topologically trivial. However, it is likely that general relativity will receive corrections
at large curvatures and the modified field equations may admit solutions corresponding to this type of geon. If geons are produced in the early universe and survive until today they could account for some of the dark matter that has been ``observed" in galaxies and galactic clusters.

In this paper I consider gravitational geons in 1+1 dimensional theories of gravity. I show that the Jackiw-Teitelboim theory
with corrections proportional to $R^2$ and $\Box R$ admits gravitational geons. I also show that gravitational
geons exist in a class of theories that includes Lagrangians proportional to $R^{2/3}$.
\newpage
\section{Introduction}
A gravitational geon is a nonsingular configuration of the gravitational field, without horizons, that persists for a long period of time \cite{Wh1,Br1} (see also \cite{3}). An interesting class of geons consists of nonsingular, asymptotically flat, topologically trivial vacuum spacetimes without horizons. It has been shown \cite{Se1,Ei1,Li1} that general
relativity does not admit such geons.
However, it is widely believed that general relativity will receive corrections in regions of large spacetime curvature. These corrections may be
either classical and/or quantum mechanical in nature (both types occur in string theory). Quantum effects are expected
to become important by the time the Planck scale is reached, but it is possible that classical corrections may appear
long before the Planck scale. It is also possible that the modified classical field equations admit gravitational geons of the type discussed above. Such geons should have masses and sizes of order of the scale at which the corrections become important. If geons are produced in the early universe and survive until today they could account for some of the dark matter that has been ``observed" in galaxies and galactic clusters (this possibility has also been discussed by Sones \cite{3} for
quantum geons with a Klein-Gordon field).

Static spherically symmetric gravitational geons have been found \cite{Dy1,Dy2} in 3+1 dimensional theories with a cosmological constant that depends on the radial coordinate and is different in the radial and angular directions.
However, it is difficult to find exact solutions to most generalizations of Einstein's equations
in 3+1 dimensions. To simplify the problem I will look for geons in two types of 1+1 dimensional theories of gravity.
The first theory that I will consider is a modified Jackiw-Teitelboim theory \cite{Ja1,Te1}. The modifications
involve adding terms proportional to $R^2$ and $\Box R$ to the field equations. Such terms, involving higher order polynomials and derivatives of the curvature are expected to occur in quantum theories of gravity.
It is shown that this theory admits gravitational geons. The second theory considered is based on the Lagrangian
\begin{equation}
L=\sqrt{-g}\left[\frac{1}{\phi}R+V(\phi)\right]\; ,
\end{equation}
which has been shown to admit nonsingular black holes, for a particular potential \cite{Tr1}. Here I show that the theory also
admits geons for certain choices of $V(\phi)$. For a particular choice of $V(\phi)$ I also show that $\phi$ can be eliminated from the action and the Lagrangian, written in terms of the Ricci scalar, is proportional to $R^{2/3}$.
\section{Field Equations and Geons}
First consider the 1+1 dimensional theory proposed by Jackiw \cite{Ja1} and Teitelboim \cite{Te1}
\begin{equation}
R=\Lambda+8\pi GT\; ,
\label{Einstein}
\end{equation}
where $R$ is the Ricci scalar, $\Lambda$ is a cosmological constant, $G$ is Newton's constant and $T$ is the
trace of the energy momentum tensor. It has been shown \cite{Ma1} that this theory possess many features in common
with 3+1 dimensional general relativity. At first sight this theory does not seem to follow from the Einstein
field equations, $G_{\mu\nu}=\lambda g_{\mu\nu}+8\pi T_{\mu\nu},$ due to the fact that $G_{\mu\nu}$ is
identically equal to zero in 1+1 dimensions. However, in 1+1 dimensions there exists a conformal anomaly
$<T>\;\propto R$ and Sanchez \cite{Sa1} has used this to show that (\ref{Einstein}) does follow from
Einstein's equations.

Now, as discussed earlier, it is believed that there will be corrections to the Einstein field equations in regions of
large spacetime curvature. These corrections may be expected to contain higher powers and derivatives of the
Riemann tensor and its contractions. In 1+1 dimensions the Riemann and Ricci tensors can be written in terms of
the Ricci scalar, so the corrections will involve higher powers and derivatives of the Ricci scalar. Here I consider
modifications to the Jackiw-Teitelboim theory of the form
\begin{equation}
R+\alpha R^2+\beta\Box R-\Lambda=8\pi GT\;,
\end{equation}
where $\alpha$ and $\beta$ are constants and $\Box=\nabla_{\mu}\nabla^{\mu}$.
These are the most general corrections involving polynomials and derivatives of $R$ that involve constants with
dimensions of (length)$^2$. For the remained of this paper I will set $\Lambda=0$ for simplicity.

In vacuum spacetimes with
\begin{equation}
ds^2=-f(r)dt^2+f^{-1}(r)dr^2\; ,
\label {metric}
\end{equation}
the Ricci scalar is $R=-f^{''}$ and the field equation is given by
\begin{equation}
f^{''}-\alpha\left( f^{``}\right)^2+\beta\frac{d}{dr}\left(ff^{'''}\right)=0\; .
\label{JTgen}
\end{equation}
Here I take $r$ to be a radial like coordinate, so that $r\geq 0$.
The general vacuum solution of the original Jackiw-Teitelboim theory (i.e. $\alpha=\beta=0$)
with $\Lambda=0$ is
\begin{equation}
f_0(r)=ar+b\; ,
\end{equation}
where $a$ and $b$ are constants. Now consider solutions to (\ref{JTgen}) of the form
\begin{equation}
f(r)=f_0(r)+\epsilon (r)
\label{sol}
\end{equation}
with $|\epsilon(r)|<<|f_0(r)|$.

First consider solutions with $a=0$ and $b=1$. The linearization of (\ref{JTgen}) gives
\begin{equation}
\epsilon^{''}+\beta\epsilon^{''''}=0\; .
\end{equation}
If $\beta<0$ set $\sigma=1/\sqrt{-\beta}$ and the general solution is
\begin{equation}
\epsilon(r)=A_1+A_2r+B_1e^{-\sigma r}+B_2e^{\sigma r}\; .
\end{equation}
Imposing the condition $|\epsilon|<<1$ gives $A_2=B_2=0$, $|B_1|<<1$ and we
can set $A_1=0$ since it just modifies the constant term in $f$. In order to be able to neglect the
$\alpha(\epsilon^{''})^2$ term relative to the $\epsilon^{''}$ term the additional constraint
\begin{equation}
\left|\frac{\alpha}{\beta}A\right|<<1
\end{equation}
must be satisfied. If $\alpha\simeq\beta$ this reduces to $|A|<<1$.
Thus,
\begin{equation}
f(r)=1+Ae^{-\sigma r}\; ,\;\;\;\;\;\;\;\; |A|<<1 \;\;\;\;and \;\;\;\;\;\left|\frac{\alpha}{\beta}A\right|<<1
\end{equation}
is an approximate solution to the field equations if $\beta<0$. In fact, if $\alpha=2\beta$ this
is an exact solution for arbitrary $A$. The Ricci scalar is given by
\begin{equation}
R=\frac{A}{\beta}e^{-\sigma r}
\end{equation}
and is bounded everywhere. Note that $R$ does not have to be small at the origin even when $|A|<<1$. For example, if $|A|\simeq 10^{-3}$ then
$R$ is about three orders of magnitute less than the scale set by $\beta$, which could be quite large.
For $|A|<<1$ there are no horizons and this solution then describes a static gravitational geon.
If $\alpha=2\beta$ we require that $A>-1$ to avoid the presence of a horizon.

If $\beta>0$ set $\sigma=1/\sqrt{\beta}$ and the relevant solution is
\begin{equation}
\epsilon (r)=A\cos(\sigma r+B)\; ,\; \;\;\;\;\;\;\;\; |A|<<1 \;\;\;\;and \;\;\;\;\;\left|\frac{\alpha}{\beta}A\right|<<1,
\end{equation}
where $A$ and $B$ are constants. This solution can be thought of as an infinite sequence of geons.

Now consider solutions of the form (\ref{sol}) with $a=1$ and $b=0$. The linearized field equation is
\begin{equation}
\beta r\epsilon^{''''}+\beta\epsilon^{'''}+\epsilon^{''}=0\; .
\end{equation}
First consider $\beta>0$ with $\sigma=1/\sqrt{\beta}$ and let $x=2\sigma\sqrt{r}$ and $y=\epsilon^{''}$.
The field equation is given by
\begin{equation}
x^2y^{''}+xy^{'}+x^2y=0\; ,
\end{equation}
which is a Bessel equation of order zero. The solution, which is nonsingular at the origin, is
\begin{equation}
y(x)=\tilde{A}J_0(x)\; ,
\end{equation}
where $\tilde{A}$ is a constant and $J_0(x)$ is the Bessel function of the first kind of order zero. This can be written as
\begin{equation}
\epsilon^{''}(r)=\tilde{A}J_0\left(2\sigma\sqrt{r}\right)
\end{equation}
and can be integrated twice using
\begin{equation}
\int x^nJ_{n-1}(x)dx=x^nJ_n(x)
\end{equation}
to obtain
\begin{equation}
\epsilon (r)=ArJ_2\left(2\sigma\sqrt{r}\right)\; ,
\end{equation}
where $A=\beta\tilde{A}$ and integration constants have been set to zero.
The function $f(r)$ can then be written as
\begin{equation}
f(r)=r\left[1+AJ_2\left(2\sigma\sqrt{r}\right)\right]\; ,\; \;\;\;\;\;\;\;\; |A|<<1 \;\;\;\;and
\;\;\;\;\;\left|\frac{\alpha}{\beta}A\right|<<1
\end{equation}
where I have imposed $|\epsilon|<<1$ and $|\alpha(\epsilon^{''})^2|<<|\epsilon^{''}|$. Since
\begin{equation}
J_n(x)\simeq\sqrt{\frac{2}{\pi x}}\cos\left(x-\frac{n\pi}{2}-\frac{\pi}{4}\right)\;\;\;\;\;\;\;\;\; as\;\;\;\;x\rightarrow\infty
\label{Bessel}
\end{equation}
we see that $f(r)\rightarrow r$ as $r\rightarrow\infty$.
The Ricci scalar is given by
\begin{equation}
R=-\frac{A}{\beta}J_0\left(2\sigma\sqrt{r}\right)
\end{equation}
and is therefore bounded and goes to zero as $r\rightarrow\infty$.

Now consider $\beta<0$ with $\sigma=1/\sqrt{-\beta}$ and define $x=2\sigma\sqrt{r}$ and $y=\epsilon^{''}$. The field equation becomes
\begin{equation}
x^2y^{''}+xy^{'}-x^2y=0\; .
\end{equation}
This is a modified Bessel equation of order zero. The general solution is
\begin{equation}
y(x)=AI_0(x)+BK_0(x)\; ,
\end{equation}
where $A$ and $B$ are constants, $I_0$ is the modified Bessel function of the first kind of order zero
and $K_0$ is the modified Bessel function of the second kind of order zero. Unfortunately, neither of these
functions is bounded on $x>0$ and we do not find geons in the linearized approximation for $\beta<0$.

Next I will consider geons in gravitational theories based on the Lagrangian
\begin{equation}
L=\sqrt{-g}\left[\frac{1}{\phi}R+V(\phi)\right]\; ,
\label{Lag1}
\end{equation}
which has been used in \cite{Tr1} to produce nonsingular black holes in 1+1 dimensions.
The field equations that follow from this Lagrangian and the metric (\ref{metric}) are \cite{Tr1}
\begin{equation}
\phi^3V(\phi)+2f\phi\phi^{''}-4f\left(\phi^{'}\right)^2+f^{'}\phi\phi^{'}=0\; ,
\label{eq1}
\end{equation}
\begin{equation}
\phi^2V(\phi)+f^{'}\phi^{'}=0\; ,
\label{eq2}
\end{equation}
and
\begin{equation}
\frac{dV(\phi)}{d\phi}=-\frac{1}{\phi^2}f^{''}\; .
\label{eq3}
\end{equation}
Solving for $V(\phi)$ in (\ref{eq2}) and substituting it into (\ref{eq1}) gives
\begin{equation}
\phi\phi^{''}-2\left(\phi^{'}\right)^2=0\; .
\end{equation}
The solution to this equation is
\begin{equation}
\phi(r)=\frac{1}{Ar+B}\; ,
\end{equation}
where $A$ and $B$ are constants. The constant $B$ can be eliminated by redefining $r$ giving
\begin{equation}
\phi(r)=\frac{1}{Ar}\; ,
\label{phi}
\end{equation}
The field equations can now be taken to be (\ref{eq2}) and (\ref{eq3}) with $\phi$ given by (\ref{phi}).
Substituting (\ref{phi}) into the two field equations gives
\begin{equation}
V(\phi)=Af^{'}\;\;\;\;\;\;\;\; and \;\;\;\;\;\;\;\; \frac{dV(\phi)}{d\phi}=-A^2r^2f^{''}\; .
\end{equation}
It is easy to show that the first equation along with $\phi=1/Ar$ implies the second equation. The remaining field equation is then given by
\begin{equation}
Af^{'}=V(\phi)\;\;\;\;\;\;\;\; with \;\;\;\;\;\;\;\;\; \phi=\frac{1}{Ar}\; .
\end{equation}
Thus, given a function $f(r)$ it is easy to solve for $V(\phi)$.

For example if
\begin{equation}
f(r)=1-\frac{2m}{r+\ell}\; ,
\label{f}
\end{equation}
where $m$ and $\ell$ are positive constants, then
\begin{equation}
V(\phi)=\frac{2mA^3\phi^2}{(1+A\ell\phi)^2}\; .
\label{potential}
\end{equation}
The Ricci scalar is given by
\begin{equation}
R=\frac{4m}{(r+\ell)^3}
\end{equation}
and is finite for all $r\geq 0$. If $\ell >2m$ there are no horizons and the solution is a static gravitational
geon. It is easy to see that other theories with different potentials can be constructed that admit gravitational geons. One simply
chooses a function $f(r)$ that describes a gravitational geon and solves for $V(\phi)$.

It is interesting to note that $\phi$ can be eliminated in favor of $R$ in the action. From (\ref{eq3}) we see that
$R=\phi^2dV/d\phi$. Using this and (\ref{potential}) it is easy to show that
\begin{equation}
A\phi=\frac{x}{1-\ell x}
\end{equation}
where
\begin{equation}
x=\left(R/4m\right)^{1/3}\; .
\end{equation}
Substituting this into the Lagrangian (\ref{Lag1}) gives
\begin{equation}
L=\sqrt{-g}\left[6mA\left(\frac{R}{4m}\right)^{2/3}-A\ell R\right]\; .
\label{Lag2}
\end{equation}
The last term can be dropped, since $R$ does not contribute to the field equations in 1+1 dimensions. The
vacuum field equations that follow from (\ref{Lag2}) and $R_{\mu\nu}=1/2Rg_{\mu\nu}$ are
\begin{equation}
Rg_{\mu\nu}=4R^{1/3}\left[g_{\mu\nu}\Box R^{-1/3}-\nabla_{\mu}\nabla_{\nu}R^{-1/3}\right]
\end{equation}
and it is easy to show that (\ref{f}) is a solution.
\section{Conclusion}
In this paper I examined two gravitational theories in 1+1 dimensions to see if they admit geons.
The first theory was a modification of the Jackiw-Teitelboim theory of the form
\begin{equation}
R+\alpha R^2+\beta\Box R-\Lambda=8\pi GT\;,
\end{equation}
where $\alpha$ and $\beta$ are constants. I showed that there are vacuum solutions with $\Lambda=0$ describing
gravitational geons.

I also examined theories that follow from the Lagrangian
\begin{equation}
L=\sqrt{-g}\left[\frac{1}{\phi}R+V(\phi)\right]\; .
\end{equation}
I showed that there exist potentials that lead to theories that admit gravitational geon solutions.
For a particular choice of $V(\phi)$ I also showed that $\phi$ can be eliminated from the action and the Lagrangian, written in terms of the Ricci scalar, is proportional to $R^{2/3}$.
\section*{Acknowledgements}
This research was supported by the  Natural Sciences and Engineering Research
Council of Canada.

\end{document}